\documentclass[twocolumn,showpacs,preprintnumbers,nofootinbib,amsmath,amssymb]{revtex4}
\usepackage{amsmath,amssymb,amsfonts}
\usepackage{graphics}
\usepackage{epsfig}
\usepackage{inputenc}
\usepackage[english]{babel}

\newcommand\al{\alpha}
\newcommand\tal{{\tilde{\alpha}}}
\newcommand\be{\beta}
\newcommand\de{\delta}

\newcommand\ga{\gamma}

\newcommand\vak{\varkappa}

\newcommand\De{\Delta}
\newcommand\Ga{\Gamma}

\newcommand\lam{\lambda}

\newcommand\half{{\frac{1}{2}}}

\newcommand\ce{{\cal E}}

\newcommand\ep{\epsilon}
\newcommand\vep{\varepsilon}
\newcommand\MD{\mathfrak{D}}
\newcommand\BMD{\bar{\mathfrak{D}}}
\newcommand\rmd{{\rm d}}
\newcommand\rmi{{\rm i}}

\newcommand\mfq{{\mathfrak q}}

\newcommand\mfp{{\mathfrak p}}

\newcommand\quat{{\frac14}}
\begin{document}
\title{New general approach in few-body
scattering calculations:\\ Solving discretized Faddeev equations on
a graphics processing unit}
\author{V.N. Pomerantsev}
\email{pomeran@nucl-th.sinp.msu.ru}
\author{V.I. Kukulin}
\email{kukulin@nucl-th.sinp.msu.ru}
\author{O.A. Rubtsova}
\email{rubtsova@nucl-th.sinp.msu.ru}
 \affiliation{Institute of
 Nuclear Physics, Moscow State University, Leninskie gory, Moscow,
119991, Russia}

\begin{abstract}
\begin{description}

\item[Background:] The numerical solution of few-body scattering
  problems with realistic interactions is a difficult problem that
  normally must be solved on powerful supercomputers, taking a lot
  of computer time.  This strongly limits the possibility of accurate
  treatments for many important few-particle problems in different
  branches of quantum physics.

\item[Purpose:] To develop a new general highly effective approach
  for the practical solution of few-body scattering equations that 
  can be implemented on a graphics processing unit.

\item[Methods:] The general approach is realized in three steps: (i)
  the reformulation of the scattering equations using a convenient
  analytical form for the channel resolvent operator; (ii) a complete
  few-body continuum discretization and projection of all operators
  and wave functions onto a $L_2$ basis constructed from
  stationary wave packets and (iii) the ultra-fast solution of the
  resulting matrix equations using graphics processor.

\item[Results:] The whole approach is illustrated by a calculation of
  the neutron-deuteron elastic scattering cross section below and
  above the three-body breakup threshold with a realistic $NN$
  potential which is performed on a standard PC using a graphics
  processor with an extremely short runtime.

\item[Conclusions:]The general technique proposed in this paper opens a
  new way for a fast practical solution of quantum few-body scattering
  problems both in non-relativistic and relativistic formulations in
  hadronic, nuclear and atomic physics.

\end{description}
\end{abstract}

\maketitle
\section{Introduction.}

It is well known that a sharp contrast exists today in the
quantum-mechanical treatment of few- and many-body systems between
very effective and fast bound-state calculations, on the one hand, and
very time-consuming few-particle scattering calculations on the other
hand. Practical solutions for a discrete spectrum may incorporate many
hundreds or even thousands particles with simple Coulomb-like
interactions in atomic or molecular physics, or up to 20-25 nucleons
with complicated realistic $NN$ interactions in nuclear physics, while
even the solution of the four-nucleon scattering problem with
realistic $NN$-interactions, especially above the three-body breakup
threshold, represent a strong challenge for modern
theorists~\cite{Lazauskas_rep}. There are at least two reasons for
such a strong contrast. First, the few-body scattering problem
includes complicated boundary conditions, especially above the three- or
four-body breakup thresholds, and second, the multi-particle
Hamiltonian has a degenerate continuous spectrum, so that each pair or
triple of particles can be in infinite number of states at the same
energy.

The first problem has been solved mathematically by  formulation of
the Faddeev--Yakubovsky equations whose full solution satisfies, as
has been strictly proved \cite{Faddeev}, all of the necessary boundary
conditions.  However, the price for this correct formulation is a
very sophisticated form of these integral equations, whose kernels
have complicated moving singularities. Therefore, in the previous
four decades a lot of  exact and approximate methods for solving
the Faddeev--Yakubovsky equations were proposed.  However, due to the
complexity of realistic few-body scattering problems, 
practical solutions usually require a massively parallel
implementation, so that even now exact Faddeev-like scattering
calculations are performed mainly on  powerful supercomputers (see
e.g. the recent $3N$-calculations \cite{gloeckle12}).  The
complexity of the few-body equations leads to the fact that an
accurate numerical treatment for realistic few-body scattering
problems remains available only to a limited number of experts.

The most effective way to treat the second key problem related to
degeneracy of few-body continuous spectra of the total and channel
Hamiltonians is their discretization by one or another method and
usage of $L_2$ normalized states as approximations for exact continuum
states.  Nowadays, many such discretization methods exist (see, for
example, a recent comprehensive review~\cite{Lazauskas_rep}). However,
it is still not clear whether these particular discretization methods
give a discrete form of scattering equations, which permit a high
degree of parallelism in a numerical solution. The last point is
crucially important for further progress in few-body scattering
calculations because even powerful supercomputers can not give
essential acceleration of the calculations if the solution method does
not permit an effective parallelization for all parts of the
algorithm.

Quite recently, a new computational technique has been introduced
based on the general-purpose graphics processing unit (GPGPU). This
technique utilizes a graphics processing unit (GPU) which has been
initially designed to carry out computations for computer
graphics. Nowadays, GPUs are specialized to perform ultra-fast general
purpose computations and they can replace a supercomputer realization
in many particular cases.  Also, special extensions of standard
programming languages are developed to use GPU facilities in tedious
scientific calculations (see e.g. \cite{CUDA}).

This technique has been actively pursued and successfully used in
quantum chemistry \cite{CUDA_ch}, in lattice QCD calculations
\cite{CUDA_QCD}, Monte-Carlo simulations etc.  Recently, {\em
  ab-initio} nuclear structure GPU-calculations \cite{Vary} have been
performed as well as GPU treatments of Faddeev equations for quantum
trimer systems \cite{yarevsky}. It is of great interest to apply such
GPU-techniques to realistic few-body scattering calculations. This
would open new possibilities for accurate few-body studies in general
and could make them more accessible to a wider number of researchers.
However, the GPU realization requires an appropriate and specific
formulation for scattering problems because this realization is most
effective for the algorithms with a high degree of parallelism and
minimal interdependence between data processing in parallel threads.

The present authors suggested in previous years the wave-packet
continuum discretization technique \cite{nd1,nd2} which has been
tested carefully for the model $NN$ interactions and found to be very
efficient.  One of the important features of the above
discrete approach is that the resulting discretized form of the scattering
equations is well suited for such a parallel realization.  Below we
show that such a massively parallel implementation of the whole
solution can be made on a standard PC with a modern graphics
processor that can perform all of the calculations using many thousands
of parallel threads.

In the present paper we have also made a generalization of the
wave-packet approach to few-body equations with fully realistic $NN$
interaction which is not a trivial problem and requires a new
determination for the multi-channel resolvent in an analytical
form. So we included a section with this
description in the present paper .

The paper is organized as follows. In Sect.~II we summarize the main
features of the WP approach and describe how it is used to solve
three-body scattering problem. In Sect.~III the case of
the coupled-channel two-body input interaction is discussed and formulas
for the three-body channel resolvent are given.  The results for the
$nd$ elastic scattering problem are represented in Sect.~IV.  Section
V is dedicated to a description of our first GPU tests for the problem
in question and a comparison of the corresponding computational efficiency 
of the CPU and GPU realizations on the same PC.  The main
results of the paper are summarized in the conclusion.

\section{New general approach in few-body scattering calculations. }

The present work discusses the solution few- and many-body scattering
problems in atomic, molecular, nuclear and hadronic physics.  Here we
discuss in detail all of the steps needed to implement our new approach
and we illustrate the whole technique using a non-trivial example ---
the solution of the Faddeev equations for $n-d$ scattering below and
above the three-body breakup threshold with a realistic $NN$
interaction.

\subsection{The basic features of the approach}
In our approach, we change all of the steps used in the conventional
procedure for solving the Faddeev equations in momentum space.

(i) The first step is to replace the conventional form of the
Faddeev integral equation, e.g. for a transition operator $U$
describing elastic $nd$ scattering \cite{gloeckle},
 \begin{equation}
 \label{Ueq0}
 U=PG_0^{-1}+Pt_1G_0U,
 \end{equation}
with the half-shell equivalent form
\begin{equation}
\label{Ueq}
 U=Pv_1+Pv_1G_1U.
 \end{equation}
 Here $v_1$ is the two-body interaction, $t_1$ is the two-body
 $t$-matrix, $G_0$ is the resolvent of the free three-body
 Hamiltonian, $H_0$, $P=P_{12}P_{23}+P_{13}P_{23}$ is the particle
 permutation operator and $G_1=[E+i0-H_1]^{-1}$ is the resolvent of
 the channel Hamiltonian
 \begin{equation}
 H_1=h_1\oplus h_0^1,
 \end{equation}
where $h_1=h_0+v_1$ is the two-body $NN$ sub-Hamiltonian and $h_0^1$ is
sub-Hamiltonian describing free motion of the third nucleon
relative to the $NN$ subsystem.  The index $1$ is
the Jacobi-set index of the initial state.

One of the main purposes for such a replacement is to change the required
two-body input: instead of fully off-shell two-body $t$-matrices at many
energies, we suggest employing two-body interactions $v_1$ in
combination with the channel resolvent $G_1$. However, in
such a replacement, one  has to evaluate additionally the channel 
resolvent operator 
$G_1$.  Fortunately, in the wave-packet approach, the finite-dimensional
approximation for this operator is calculated easily in a closed 
analytical form \cite{nd1}.

Moreover, the whole energy dependence appears now in the channel
resolvent operator (rather than in the off-shell $t$-matrix, as in the
conventional formulation), which is calculated explicitly. So
that, with such a replacement, we can find a solution of few-body
scattering equations {\em at many energies almost with the same
computational effort that is needed for a single energy}.

Another important advantage of our approach is a new treatment
of three-body breakup. Contrary to the conventional approach, we
treat  three- or  many-body breakup processes as  particular cases
of inelastic excitations (into states of the discretized continuum)
\cite{nd2}. Such a treatment strongly facilitates breakup
calculations.

(ii) The second step is to project the integral kernels of the
reformulated Faddeev equations and the solution onto a special
orthogonal basis of the stationary wave packets (WPs), which
corresponds to a formulation of the scattering problem on a momentum
lattice.  Such basis is very appropriate for constructing normalized
analogs of continuum states for the channel Hamiltonian $H_1$. It
follows that the solution of the three-body scattering problem is
described in the terms of asymptotic channel states, in contrast to
conventional approach which employs free three-body states (the plane
waves).

Such a projection of the three-body scattering equations onto a
three-body WP basis results in matrix equations which allow us to
circumvent the main difficulties that arise in the conventional
solution of the initial singular integral equation. Firstly, the use
of a finite matrix for the permutation operator in a discrete WP basis
eliminates the need for the very numerous multi-dimensional
interpolations of a given solution into the ``rotated'' Jacobi set
during iterations. Further, all singularities of the Faddeev kernel
(in the form $Pv_1G_1$) are isolated now in the channel resolvent $
G_1 $ and thus can be easily smoothed and averaged when using the WP
representation \cite{nd1}.  At last, the resulting matrix equations
can be solved directly at real energies without any contour rotations
or deformations onto complex plane, which are often employed in the
solution of singular integral equations.

The resulting matrix equation (of high dimension) obtained in the WP
approach is solved by simple iterations when they converge or
otherwise by applying an additional Pade-approximant summation. The
computational scheme turns out to be very efficient and thus the
whole calculation can be performed even on a standard PC.

Following these steps, in our previous papers, \cite{nd1,nd2}, we
studied elastic scattering and breakup cross sections in a $3N$ system
with a central $NN$ potential. But it was still unclear if the
advantages of the above computational scheme remain valid for
realistic $NN$ interactions including tensor, spin-orbit
etc. components and in the particularly when the number of contributing 
spin-orbital partial channels is large\footnote{In particular, the authors of the recent review
  \cite{Lazauskas_rep} expressed some doubts in full applicability of
  the present WP-approach to realistic interactions.}. 
To investigate this question we will apply our approach to a 
three-nucleon system interacting with a realistic $NN$ interaction, including
a tensor component (the Nijmegen $NN$ potential \cite{nijm}), at
energies below and above three-body breakup threshold.

(iii) To further extend the complexity of scattering problems that can
be treated accurately, it is desirable to develop a highly parallel
algorithm for the solution of the resulting matrix equations of large
dimension. {\em The third step} is to parallelize the
algorithm to adapt it for computations by a GPU.  Such a GPU realization
is shown in the present paper to make the solution of resulting matrix
equations (derived from multi-channel system of integral Faddeev
equations) extremely fast even on a standard PC.

\subsection{Discrete  form of the Faddeev equation in  wave-packet
representation.}

Here we briefly describe our approach based on a continuum
discretization using the stationary wave packets.  
We illustrate this using the example of
the Faddeev equation (\ref{Ueq}) for the transition operator $U$ for
$nd$ scattering (further details see in \cite{nd1,nd2}).

 \subsubsection{Definition of momentum lattice basis functions}
 
 To construct the three-body WP basis functions, we start from the two-body
 case and introduce partitions of the continua of two free
 sub-Hamiltonians, $h_0$ and $h_0^1$, onto non-overlapping intervals
 $\{\mathfrak{D}_i\equiv[\ep_{i-1},\ep_i]\}_{i=1}^M$ and
 $\{\bar{\mathfrak{D}}_j\equiv[\ce_{j-1},\ce_j]\}_{j=1}^N$
 respectively. These sub-Hamiltonians describe the free motion of
 particles 2 and 3 with relative momentum $p$ and the free motion of
 particle 1 with momentum $q$ relative to the center of mass of the
 pair (23), respectively.  Thus the free stationary wave packets
 $|\mathfrak{p}_i\rangle$ and $|\mathfrak{q}_j\rangle$ are built as
 integrals of free solutions $|p\rangle$ and $|q\rangle$ over the
 discretization bins:

\begin{equation}
\label{pq}
|\mathfrak{p}_i\rangle=\frac{1}{\sqrt{B_i}}\int_{\mathfrak{D}_i}f(p)|p\rangle
dp,\
|\mathfrak{q}_j\rangle=\frac{1}{\sqrt{\bar{B}_j}}\int_{\bar{\mathfrak{D}}_j}\bar{f}(q)|q\rangle
dq,
\end{equation}
where $B_i,\bar{B}_j$ and $f(p),\bar{f}(q)$  are  normalization
factors and weight functions respectively \cite{nd1,nd2}. Here and
below we denote the functions and values corresponding to
$q$-variable with additional bar mark to distinguish them from the
functions corresponding to the $p$-variable.

When constructing the three-body WP basis one should take into
account  spin and angular parts of the basis functions. We use the
following quantum numbers for the subsystems and the whole
three-body system  according to the $(jj)$-coupling scheme:
\begin{equation}
\alpha=\{l,s,j\};\qquad \beta=\{\lambda,I\}; \qquad
\Gamma=\{J,\pi,T\}, \label{qnum}
\end{equation}
where  $l,s$ and $j$ are  $NN$ quantum numbers: $l$ is an orbital
momentum, $s$ is a spin and ${\bf j}={\bf l}+{\bf s}$ is a total
angular momentum of the  subsystem (the interaction potential
depends on the value of $j$). The other quantum numbers are the
following:  $\lambda$ is an orbital momentum   and ${\bf
I}=\mbox{\boldmath$\lambda$}+\mbox{\boldmath$\sigma$}$ is a total
momentum of the third nucleon, where $\sigma=\half$ is its spin.
Finally, ${\bf J}={\bf j}+{\bf I}$ is a total angular momentum of
the three-body system, $T$ is the total isospin  and $\pi$ is parity,
all of them are conserved. Let's also note that the pair
isospin $t$ can be defined by values of $l$ and $s$, because the sum
$l+s+t$ must be odd.

The free WP states should be defined for each partial wave $l$ and
$\lam$ and further they are multiplied by appropriate spin-angular
states.  Thus the three-body basis function can be written as:
\begin{equation}
\label{xij}
|X_{ij}^{\Ga\al\be}\rangle=|\mathfrak{p}_i^l\rangle\otimes|\mathfrak{q}_j^\lam\rangle
|\alpha,\beta:\Gamma\rangle,
\end{equation}
where $|\alpha\rangle$ is a spin-angular state of the $NN$ pair,
$|\beta\rangle$  is a  spin-angular state of the third nucleon,
and $|\Gamma\rangle$ is a set of  three-body quantum numbers.

The state (\ref{xij}) is the WP analog
of the exact plane wave state in three-body continuum
$|p,q;\alpha,\beta:\Gamma\rangle$ for the three-body free Hamiltonian
$H_0=h_0\oplus h_0^1$.

The  free stationary wave packets defined in eq.~(\ref{pq}) with
unit weights are step-like functions in the momentum representation
\cite{nd1,nd2} while the  three-body free WP basis functions are
constant inside the cells of the lattice built by a convolution of
two one-dimensional cells $\{\mathfrak{D}_i\}_{i=1}^M$ and
$\{\bar{\mathfrak{D}}_j\}_{j=1}^N$. We refer to the free WP
basis as {\em a lattice} basis.  We denote the two-dimensional
bins (i.e. the lattice cells) by $\MD_{ij}=\MD_i\otimes\BMD_j$.

\subsubsection{The  wave-packet basis for the channel Hamiltonian}

In the case of a single-channel two-body input interaction (e.g. the
central one), we have demonstrated \cite{nd1,nd2} that it is
possible to define {\em scattering WPs} corresponding to the exact
scattering wave functions $|\psi_p\rangle$ of the sub-Hamiltonian
$h_1$:
\begin{equation}
\label{z_k}
|z_k\rangle=\frac{1}{\sqrt{C_i}}\int_{{\De}_i}w(p)|\psi_p\rangle dp,
\end{equation}
where $\De_i$ are partition intervals and $C_i$ and $w(p)$ are a
normalization factor and a weight function.

To use the states (\ref{z_k}) practically, one can approximate them
with the pseudostates of the sub-Hamiltonian $h_1$ in some $L_2$ basis
\cite{nd1,nd2}. Also it has been shown that the free WP basis is very
appropriate to approximate scattering states because the respective
functions have a very long-range behavior in configuration space.
So we can calculate the eigenstates (the bound and pseudostates)
of the sub-Hamiltonian $h_1$ matrix in the two-body WP-basis
$\{|\mathfrak{p}_i\rangle\}_{i=1}^M$ via a diagonalization procedure.
As a result one gets the eigenstates of the $h_1$ sub-Hamiltonian expanded
in the free WP basis (for each partial wave $l$):

\begin{equation}
\label{exp_z} |z_k^l\rangle\approx\sum_{i=1}^M
O_{ki}^l|\mathfrak{p}_i^l\rangle.
\end{equation}

For the case of a central interaction, the three-body quantum numbers for the
channel Hamiltonian $H_1$ are the same as those for the three-body free
Hamiltonian $H_0$, so that total three-body WP states corresponding to
the channel Hamiltonian $H_1$ (three-body scattering wave packets ---
SWP) are built as direct products of the two-body WPs with the same
spin-angular quantum numbers $\al,\be,\Gamma$:
\begin{equation}
 |Z_{kj}^{\Ga\al\be}\rangle\equiv
|z_k^l\rangle\otimes|\mathfrak{q}_j^\lam\rangle|\alpha,\beta:\Gamma\rangle.
\label{3swp}
\end{equation}
The main advantage of the basis constructed from SWP (\ref{3swp}) is
that one gets {\em an explicit analytical} and even {\em diagonal
form} for the matrix of three-body channel resolvent $G_1$
\cite{nd1,nd2}.

Having the WP basis for the channel-Hamiltonian
(\ref{3swp}) at our disposal, 
it is possible to project all of the constituents of the
integral equation (\ref{Ueq}) and find its finite-dimensional, i.e.
matrix, analog.  As an important result of the projecting of the
channel operator $G_1v_1$ onto the SWP-basis, one gets the main part
of the Faddeev kernel matrix in a convenient analytical
form, with a completely analytical energy dependence, in
sharp contrast to a conventional approach with a fully off-shell
$t$-matrix in a numerical form.

\subsubsection{The matrix of the permutation operator}

The permutation operator matrix $\mathbb P$ in the three-body SWP basis
$|Z_{kj}^{\Ga\al\be}\rangle$ can be expressed through the overlap
matrix ${\mathbb P}^0$ in the free WP basis
$|X_{ij}^{\Ga\al\be}\rangle$ using the rotation matrices $\mathbb O$
from the expansion (\ref{exp_z}) :
\begin{equation}
 \label{perm_z}
\langle Z_{kj}^{\Ga\alpha\beta}|P|Z_{k'j'}^{\Ga\alpha'\beta'}\rangle
\approx \sum_{ii'}O_{ki}^l O_{k'i'}^{l' *}\langle
X_{ij}^{\Ga\alpha\beta}|P|X_{i'j'}^{\Ga\alpha'\beta'}\rangle.
\end{equation}

A matrix element of the operator $P$ in the free WP basis is equal
to the overlap between basis functions defined in different Jacobi
sets.
Such a matrix element can be
calculated by integration with the basis functions over the momentum
lattice cells:
\begin{eqnarray}
\langle X_{ij}^{\Ga\al\be}|P|X_{i'j'}^{\Ga\al'\be'}\rangle=
\int_{\MD_{ij}}p^2dpq^2dq\int_{\MD'_{i'j'}}(p')^2dp'(q')^2dq'\times\nonumber\\
\frac{f^*(p)\bar{f}^*(q)f(p')\bar{f}(q')}{\sqrt{B_iB_{i'}\bar{B}_j\bar{B}_{j'}}}\langle
pq,\al\be:\Ga|P|p'q',\al'\be':\Ga\rangle,\label{perm}\qquad \qquad
\end{eqnarray}
where the prime on the lattice cell $\MD'_{i'j'}$ indicates that the cell
belongs to the other Jacobi set while $\langle
pq,\al\be:\Ga|P|p'q',\al'\be':\Ga\rangle$ is the kernel of particle
permutation operator in momentum space. This kernel, as is well known
\cite{gloeckle}, is proportional to the product of a Dirac delta and
a Heaviside theta function. However, due to integration in the
eq.~(\ref{perm}), these singularities get averaged over the momentum
lattice cells and, as a result, the elements of the permutation
operator matrix in the WP basis are finite and regular.

The matrix element in the eq.~(\ref{perm}) can be calculated using a
double numerical integration. The practical technique of such
calculation is described in the Appendix to the present paper.

\subsubsection{Matrix analog of the Faddeev equation for 
elastic $nd$ scattering and breakup}

Having evaluated the matrix of permutation operator $P$, the
calculation of the kernel $Pv_1G_1$ matrix becomes fast and
straightforward.

As a result of projecting the integral equation
(\ref{Ueq}) onto the three-body SWP basis, one gets its matrix
analog (for the each set of three-body quantum numbers $\Gamma$):
\begin{equation}
\label{m_pvg} {\mathbb U}={\mathbb P}{\mathbb V}_1+{\mathbb
P}{\mathbb V}_1 {\mathbb G}_1 {\mathbb U}.
\end{equation}
Here ${\mathbb P}$, ${\mathbb V}_1$ and ${\mathbb G}_1$ are matrices
of the permutation operator, the pair interaction and the channel
resolvent respectively defined in the SWP basis\footnote{A 
similar reduction to the discrete matrix form can be done also for
Lippmann--Schwinger, Faddeev--Yakubovsky and relativistic Faddeev
equations}.

The on-shell elastic amplitude for the $nd$ scattering in the WP
representation is defined by
a diagonal  matrix element of the $\mathbb
U$-matrix \cite{nd1,nd2}:
\begin{equation}
A_{\rm el}^{\Gamma\al_0\be}(q_0)\approx  \frac{2m}{3q_0}
\frac{\langle
Z^{\Ga\alpha_0\beta}_{0j_0}|\mathbb{U}|Z^{\Ga\alpha_0\beta}_{0j_0}\rangle}{\bar{d}_{j_0}},
\end{equation}
where $m$ is the nucleon mass, $q_0$ is the initial two-body momentum
and $|Z^{\Ga\alpha_0\beta}_{0j_0}\rangle=|z_{0}^{\alpha_0},{\mathfrak
  q}_{j_0}^\lam;\al_0,\be:\Ga\rangle$ is the SWP basis state
corresponding to the initial scattering state. Here
$|z_{0}^{\al_0}\rangle$ is the bound state of the pair (the deuteron,
in our case), the index $j_0$ denotes the bin $\BMD_{j_0}$, including
the on-shell momentum $q_0$, and $\bar{d}_{j_0}$ is the momentum width of
this bin.

It should be noted here that in our discrete WP approach {\em the
three-body breakup is treated as a particular case of an inelastic
scattering} \cite{nd2} (defined by the transitions to the specific
two-body discretized continuum states), so that the breakup
amplitude  can be defined in terms of {\em the same matrix} $\mathbb
U$ determined from eq.~(\ref{m_pvg}).

For the case  of the tensor components of the $NN$ interaction (or
other coupled-channel two-body interactions $v_1$), the
generalization of the whole formalism is straightforward. However,
it is necessary to take into account some new aspects related to
the discretized spectrum of the two-body {\em  multichannel} Hamiltonian
$h_1$ to build the correct approximation for the discretized resolvent
$G_1$.

\section{Channel resolvent in case of a coupled-channel
 interaction.}

\subsection{Construction of SWP basis for a coupled-channel sub-Hamiltonian}

The three-body SWP states corresponding to the channel Hamiltonian
$H_1$ can be defined similarly to the one-channel two-body interaction
case, i.e. as direct products of two-body WP states for the $h_0^1$ and
$h_1$ sub-Hamiltonians (jointly with the bound-state) multiplied by
spin-angular functions of the system. However, here the possible
spin-angular couplings in the $\{23\}$ subsystem due to the tensor
component in a pairwise interaction $v_1$ should be taken into
account.

Recently, the present authors have developed a convenient
approach for solving multichannel scattering problems via a
straightforward  diagonalization of the multichannel Hamiltonian in
a WP basis --- the  discrete
spectral-shift  (DSS) formalism \cite{KPRF}. In this approach, the
multichannel Hamiltonian pseudostates have been shown to correspond
to scattering wave functions defined in the so called eigenchannel
representation (ER) \cite{greiner} for which the multichannel
$S$-matrix is diagonal.

Let's introduce two types of  scattering states of the
sub-Hamiltonian $h_1$: the scattering states $|\psi_p^{l}\rangle$
(which include radial and orbital parts) corresponding to the
initial wave with a definite orbital momentum $l$ and the scattering
states defined in the ER $|\psi^{\varkappa}_p\rangle$, where
$\varkappa$ is an eigenchannel index. In case of a tensor $NN$
interaction, the scattering states $|\psi^{\varkappa}_p\rangle$ are
linear combinations of states $|\psi_p^{l}\rangle$, e.g. the coupled
pairs $^3S_1-{}^3D_1$, $^3P_2-{}^3F_2$ etc.

The main advantage of the ER formalism is that
one gets the following spectral expansion for the resolvent of the
pair sub-Hamiltonian $g_1(E)\equiv [E+i0-h_1]^{-1}$ (see e.g.
\cite{KPR_Yaf14}):
\begin{equation}
\label{2g1} g_1(E)=\frac{|z_0\rangle\langle
z_0|}{E-\epsilon_0}+\sum_{\varkappa}\int_0^\infty
\frac{|\psi^{\varkappa}_{p}\rangle\langle\psi^{\varkappa}_{p}|}{E+i0-\frac{p^2}{m}}dp,
\end{equation}
which is {\em diagonal} in the eigenchannel index $\varkappa$. In
eq.~(\ref{2g1}),  $\epsilon_0$ is an energy of the bound state
$|z_0\rangle$ and $m$ is the nucleon mass.

Thus, it is convenient to construct multi-channel two-body SWP's
as integrals of exact scattering wave functions of the
sub-Hamiltonian $h_1$ defined in the ER:
\begin{equation}
\label{mult_zi} |z_{\vak,i}\rangle=
\frac{1}{\sqrt{C_{i}^{\vak}}}\int_{\De_i^{\vak}}
w(p)|{\psi}_{p}^{\vak}\rangle dp,\quad i=1,\ldots,M^{\vak},
\end{equation}
where $\De_i^{\vak}\equiv[E_{i-1}^{\vak},E_i^{\vak}]$ are new
partition intervals whose parameters might be different for
different $\vak$.

The crucial feature of the DSS formalism defined in \cite{KPRF} is
that these multichannel SWP can be constructed  (jointly with the
deuteron bound state wave function) as pseudostates of the
Hamiltonian $h_1$ matrix in the multichannel free WP basis
$|\mfp_{i}^{l}\rangle$  (where radial and angular parts of wave
functions included).  Because $l$ is not conserved, for each value
of total momentum $j$, the two-channel $h_1$ sub-Hamiltonian states
are constructed from free WP bases
$\{|\mfp_{i}^{l}\rangle\}_{l=l_1,l_2}$ defined for both possible
values of $l_1$ and $l_2$.

Finally, we have at our disposal the multichannel SWP basis functions
which are related to the free WPs by a simple orthogonal
transformation as in the one-channel case (\ref{exp_z}):
\begin{equation}
\label{rotation}
|z_{\varkappa_\tal,k},\tal\rangle=\sum_{i=1}^M\sum_{l}
O_{ki}^{\varkappa_\tal l}|\mathfrak{p}_{i}^{l},\al\rangle,
\end{equation}
where the spin-angular parts of wave functions are taken into account as
well and the multi-index $\tal=\{\vak_\tal,s,j\}$ related to the ER
is introduced.  Below we will not detail the index $\vak_\tal$ 
which is 
a part of the multi-index $\tal$.

In this way, we construct the three-body SWP basis functions for the
channel Hamiltonian $H_1$ as direct products of the two-body ones
for the $h_0^1$ and $h_1$ sub-Hamiltonians:
\begin{equation}
\label{si} |Z^{\Ga\tal\beta}_{kj}\rangle
\equiv|z_{\vak_\tal,k}^{\tal}\rangle\otimes|\mathfrak{q}_j^\beta\rangle
|\tal,\beta:\Gamma\rangle.
\end{equation}
These states are WP analogs of the three-body ER scattering states
$|\psi^{\varkappa_\tal}_{p},q;\tal,\be:\Ga\rangle$ of the channel
Hamiltonian $H_1$.

Hence, starting from free WP bases for each two-body sub-Hamiltonian
one gets a set of basis states both for the three-body free and
channel Hamiltonians, $H_0$ and $H_1$ respectively, which are
related to each other by a simple matrix rotation.

\subsection{Resolvent of the channel Hamiltonian}

The spectral expansion of the three-body channel resolvent $G_1$ in the
ER can be found straightforwardly by making a convolution of
the multi-channel subresolvent $g_1$ with the resolvent
$g_1^0(E)=[E+i0-h_0^1]^{-1}$ for the free sub-Hamiltonian $h_0^1$ of the
third nucleon

\begin{equation}
G_1(E)=\frac{1}{2{\pi\rm i}}\int_{-\infty}^{\infty}{\rm d}\vep
g_1(\vep)g_0^1(E-\vep).
\end{equation}
This leads to the following representation for the
multi-channel three-body resolvent operator $G_1$ via the scattering
eigenfunctions of the $NN$ subsystem defined in the ER:
\begin{eqnarray}
G_1(E)=\sum_{\Ga,\tal_0,\be}  \int_0^\infty\rmd q
\frac{|z_{0},q,\tal_0\be:\Ga\rangle \langle z_0,q,\tal_0\be:\Ga
|}{E+\rmi 0
-\ep_{0}-\frac{3q^2}{4m}}\nonumber+\\
\sum_{\Ga,\tal,\be} \int_0^\infty\rmd p\rmd q
\frac{|\psi^{\varkappa_\tal}_{p},q;\tal,\be:\Ga\rangle\langle
\psi^{\varkappa_\tal}_{p},q;\tal,\be:\Ga|}{E+\rmi 0
-\frac{p^2}{m}-\frac{3q^2}{4m}}.\label{g11}
\end{eqnarray}
The first term (the bound-continuum part) in eq.~(\ref{g11}) is a
spectral sum over three-body states corresponding to the free motion of
the third nucleon relative to the deuteron. The second term (the
continuum-continuum part) in eq.~(\ref{g11}) for the channel resolvent
includes channel three-body states with $NN$ pair interacting in the
continuum (in the ER) and its imaginary part is defined by a
discontinuity across the three-body cut on the Riemann energy surface.

By projecting  the  channel resolvent operator onto the three-body
SWP basis defined in eq.~(\ref{si}), one can find  analytical
formulas for  matrix elements  of operator $G_1$ in such a
basis. The respective matrix is diagonal in all wave-packet and spin
indices:
\begin{equation}
\langle Z_{kj}^{\Ga\tal\beta}|G_1(E)|
Z^{\Ga\tal'\beta'}_{k'j'}\rangle= \de_{kk'}\de_{jj'}\de_{\tal\tal'}
\de_{\beta\beta'} G^{\Gamma\tal\beta}_{kj}(E).
 \label{dg1}
\end{equation}
Here the diagonal matrix elements $G^{\Gamma\tal\beta}_{kj}(E)$
depend, in general, on the spectral partition parameters (i.e.
$\De_k^{\vak_\tal}$ and $\bar{D}_j$ values) and total energy $E$
only. These matrix elements do not depend explicitly on the
interaction potential $v_1$.

The matrix elements $G^{\Gamma\tal\beta}_{kj}(E)$  are defined as
integrals over the respective momentum intervals for the
bound-continuum part of the whole operator:
\[
G^{\Gamma\tal_0\be}_{0j}(E)=\frac1{{\bar
B}_j}\int_{\BMD_j}\frac{|\bar{f}(q)|^2{\rm d}q }{E+{\rm
i}0-\epsilon_n^{\al}-\frac{3q^2}{4m}}, \eqno(\ref{dg1}a)
\]
and for the
continuum-continuum part:
\[
G^{\Gamma\tal\be}_{kj}(E)=\frac1{C_k^{\varkappa_\tal}
\bar{B}_j}\int_{\De_{k}^{\varkappa_\tal}}\int_{\BMD_j}
\frac{|w(p)|^2|\bar{f}(q)|^2{\rm d}p{\rm d}q }{E+{\rm
i}0-\frac{p^2}{m}-\frac{3q^2}{4m}}. \eqno(\ref{dg1}b).
\]
If the solution of the scattering
equations in the finite-dimensional WP basis converges with
increasing the basis dimension,  the final result turns out to be
{\em independent} of the particular spectral partition parameters.

Representations (\ref{dg1}a) and (\ref{dg1}b) for the channel
resolvent are the basis of the wave-packet approach, since, after a
straightforward analytical evaluation of the integrals\footnote{We
  have found previously \cite{Yaf} the explicit formulas for the
  resolvent matrix elements (\ref{dg1}) when one uses the WP's with
  the weight functions $w(p)\sim\sqrt{p}$, $\bar{f}(q)\sim\sqrt{q}$.}
in (\ref{dg1}a) and (\ref{dg1}b), one gets explicit formulas for the
three-body resolvent and thus a drastic simplification of the solution
of a general three-body scattering problem. These analytical
expressions can be used directly to solve the matrix Faddeev equation
in eq.~(\ref{m_pvg}).

\section{Solution of the n-d scattering problem with a 
realistic $NN$ interaction}

In this section the effectiveness of the new approach in few-body
calculations will be illustrated by solving the Faddeev equations for
$nd$ scattering with the realistic Nijmegen $NN$
potential~\cite{nijm}.

The $S$-matrix of elastic $Nd$ scattering is usually parameterized
by the eigen phase shifts and mixing angles in the total
channel spin $\mbox{\boldmath$\Sigma$}={\mathbf
s}+\mbox{\boldmath$\sigma$}$ representation (here $s$ is the $NN$
pair spin and $\sigma$ is a spin of the third nucleon). Unitary
transformation between amplitudes defined in the representation of
the total angular momentum of the third nucleon $I$ and the
$\Sigma$-representation is given by the matrix \cite{gloeckle}:
\begin{eqnarray}
U^J_{\lambda'\Sigma',\lambda\Sigma}=\sum_{I,I'}\sqrt{\hat{I}'\hat{\Sigma}'}
\,(-1)^{J-I'} \left\{\begin{array}{ccc}\lambda'&1/2&I'\\j&J&\Sigma'
\end{array}\right\}  \times \nonumber\\
\sqrt{\hat{I}\hat{\Sigma}}\,(-1)^{J-I}\left\{\begin{array}{ccc}\lambda&1/2&I\\j&J&\Sigma
\end{array}\right\} U^J_{\lambda' I',\lambda I}.
\end{eqnarray}

\begin{figure}[h!] \epsfig{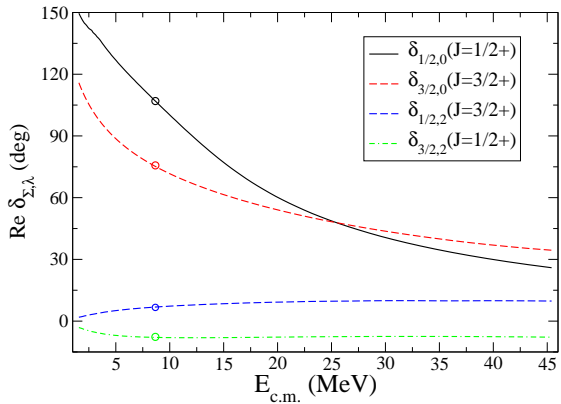}
\caption{(Color online) Some of $S$- and $D$-wave partial phase
shifts of the elastic $nd$ scattering obtained within the WP
approach (solid lines) and within the standard Faddeev calculations
 (circles)~\cite{gloeckle}.\label{phases} }
\end{figure}

\begin{figure}[h!]
\epsfig{file=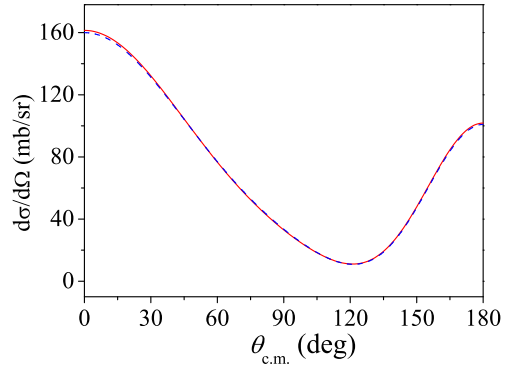,width=0.9\columnwidth} \caption{(Color online)
The differential cross section for elastic $nd$ scattering at
13~MeV obtained via the  WP technique (solid line) and within the
standard Faddeev calculations~\cite{gloeckle} (dashed line).
\label{cross13}}
\end{figure}

\begin{figure}[h!]
\epsfig{file=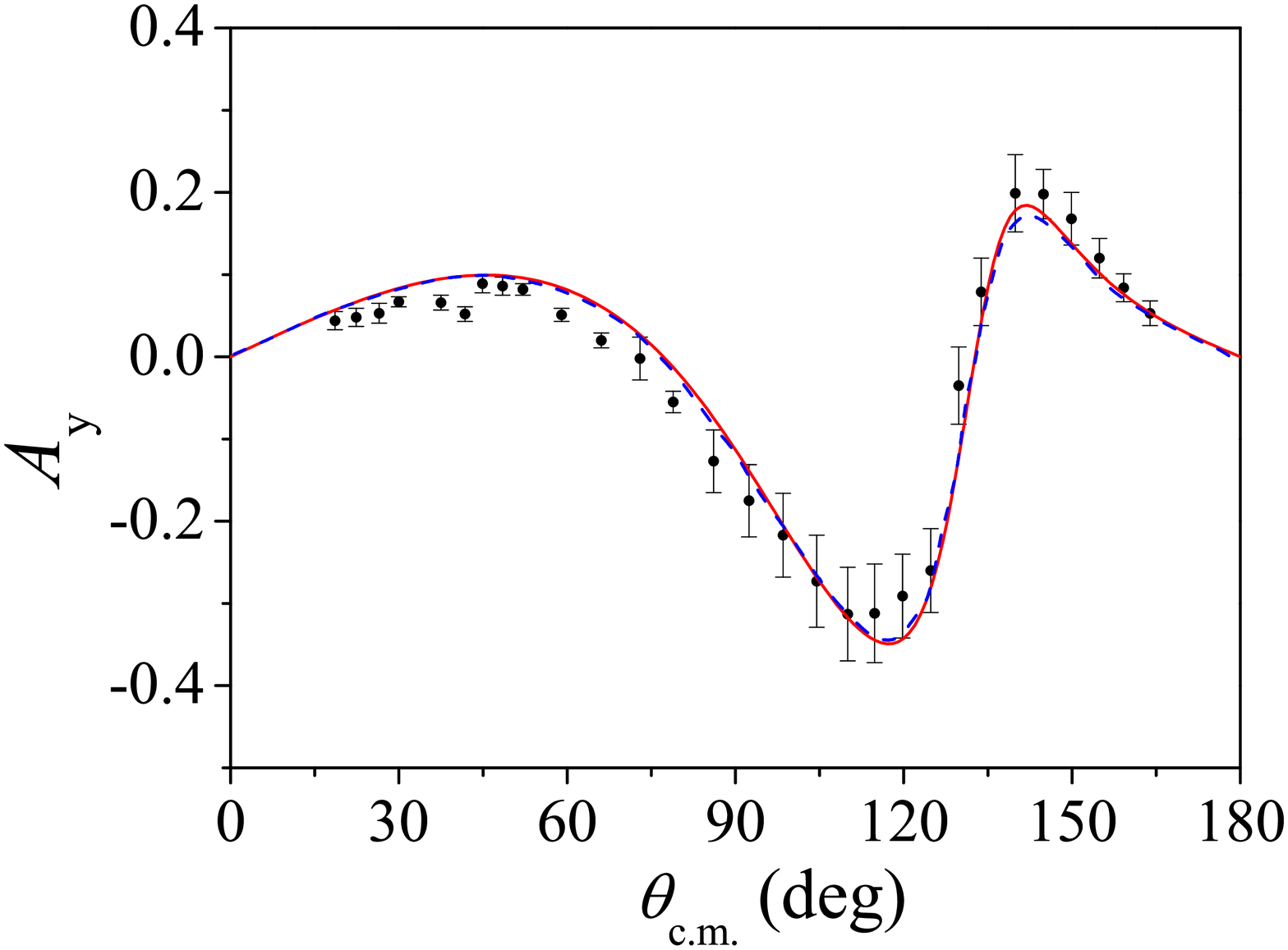,width=0.9\columnwidth} \caption{(Color online)
The neutron vector analyzing power $A_y$ for the elastic  $nd$
scattering at 35~MeV obtained within WP approach (solid line) 
compared to the result from~\cite{gloeckle} (dashed line) and
the experimental  $pd$ data \cite{pd_exp}.\label{ay} }
\end{figure}

In the Figs.~\ref{phases}-\ref{ay} the results of our fully discrete
calculations for elastic $nd$ scattering with the Nijm~I $NN$
interaction performed within the WP approach are compared with the
results of conventional Faddeev calculations of the Bochum--Krakow
group~\cite{gloeckle}.
 
In the example, we restricted ourselves to the total
isospin value $ T = 1/2 $ and took into account all of the pairwise
$NN$ channels with  two-body total angular momentum  $j\le 3$
(this gives up to 54 spin-angular channels).

In  Fig.~\ref{phases}  the results for the lowest even partial
phase shifts $\delta^{J\pi}_{\Sigma\lambda}$ of elastic $nd$
scattering both below and above a three-body breakup threshold are
shown. In the example, we employed WP bases with dimensions $M=
N=144$ along two Jacobi coordinates. This gives a matrix system
with the dimension $144\times 144=20736$  which has been handled
easily on an ordinary PC in contrast to typical conventional
calculations for similar Faddeev system which require supercomputer
facilities (see e.g. \cite{gloeckle12})\footnote{Although the matrix
dimension in our approach is much higher than in the
conventional approach, our kernel contains a very sparse permutation
matrix with only ca. 1\% of non-vanishing matrix elements. }.
It should be especially emphasized that  the
calculation of the phase shifts at 100 different energy values
displayed in  Fig.~\ref{phases} takes in our approach only about
twice as much time as the calculation for a single energy because
for all energies we employ the same permutation matrix $\mathbb P$
which is calculated only once.

In  Fig.~\ref{cross13}  our results for the differential cross
section for elastic $nd$ scattering at 13~MeV are compared to the
results of conventional Faddeev calculations\footnote{For this
comparison we  employed  the partial phase shifts and mixing angles
given in the Ref.~\cite{gloeckle} for values of the total angular
momentum at $J\le 7/2$.}~\cite{gloeckle}. While in  Fig.~\ref{ay}
the same comparison is given for  the neutron vector analyzing
powers $A_y$ for the elastic  $nd$ scattering at 35~MeV.  Here the
WP basis with dimension $M\times N=100\times 100$ has been used and
the partial waves with the total angular momentum up to $J\le 17/2$
have been taken into account.

It is clear from  all of the above illustrative examples that one can
achieve almost perfect agreement between our results and the results
of the standard Faddeev calculations \cite{gloeckle} performed on a
powerful supercomputer.
 
The present calculations were performed on the serial PC
with an Intel i7-3770K (3.50GHz) processor with 32 GB of RAM. The
real CPU computational time (which includes the permutation matrix
evaluation), i.e. the total calculation which starts from two-body
interaction potential and ends with partial three-body amplitudes
for a 54-channel calculation with the basis dimension $10^4$
for a single value of  the total angular momentum $J$ and
parity takes about 7 minutes.

Although the whole calculation for many partial waves will take a
longer time, there is the possibility within the present approach to
accelerate the whole calculation and thus solve more complicated
scattering problems already using an ordinary PC.  So, keeping in mind
our general aim to simplify and accelerate drastically realistic
many-body scattering calculations in nuclear, atomic, molecular
etc. studies by using a discrete matrix reduction of the integral
scattering equations in the WP scheme, we propose to employ the ultra-fast
GPGPU technique to further optimize the solution of the resulting
matrix equations.

\section{Solution of the Faddeev matrix equation using GPU}
 
In this section, we demonstrate a high efficiency of using a graphics
processor in numerical solution of the above matrix equation for the
three-body scattering problem.

\subsection{Parallelization of a  numerical algorithm.}

First we describe the overall numerical scheme for solving
the three-body scattering problem in the WP formalism, paying
particular attention to those aspects that are important for an
efficient parallelization.

The use of a fixed matrix for the permutation operator
completely eliminates the necessity of the numerous and time
consuming interpolations of a current solution in the iterations.
These computations take the majority of the computing time in
the standard integral approach\footnote{Note
that such very numerous multi-dimensional interpolations at every
step of iterations seem to be hardly realizable via highly  parallel
execution.}. Contrary to this, in our approach, the main
computational effort (in case of CPU-calculations) are spent on a
calculation of the permutation matrix ${\mathbb P}^0$ in the lattice
WP basis. However, the matrix ${\mathbb P}^0 $ is independent of
energy, and therefore, being calculated once, it can be used to
solve the scattering problem at many energies, as well as for
various two-particle input interactions. While in the standard
approach, the whole calculation must be repeated for each energy and
for each type of two-body interaction.

The main difficulty which is met in the practical solution of the
matrix equation (\ref{m_pvg}) is its large dimension. So, it is
impossible even to store the entire matrix of the kernel $\mathbb K
={\mathbb P}{\mathbb V}_1 {\mathbb G}_1 $ in a RAM of a computer.
However, one can effectively employ the fact that the matrix $\mathbb
K $ can be represented as a product of four matrices: a very sparse
matrix ${\mathbb P}^0$ (only about 1~\% of its matrix elements are
nonzero \cite{nd2}), a diagonal matrix $G_1$ and two block
matrices. So that, it is sufficient to store only factors of the
matrix $\mathbb K$ instead of all its matrix elements ( only the
nonzero elements of the sparse matrix are stored). In this way, the
required memory can be reduced by about 100 times. This eliminates the
need for an external memory in the process of iteration, which in turn
reduces a computational time by an order of magnitude even for a
conventional CPU realization.

So, our overall numerical scheme should be quite suitable for
parallelization and implementation on the multiprocessor systems,
particularly on a GPU. The optimized algorithm for the solution of
the realistic $nd$ scattering problem  
consists of the following main steps:\\

1. Construction of the three-body SWP basis including
 preparation of two-body bases (via diagonalization of the pairwise
$NN$ sub-Hamiltonian matrix in the free WP basis);  the  calculation
of the algebraic coefficients $g_{\ga,\ga'}$ (\ref{gal}) for the coupling of
different spin-angular channels.\\
2. Selection of the nonzero elements of the overlap matrix ${\mathbb P}^0$.\\
3. Calculation of  the nonzero elements of ${\mathbb P}^0$.\\
4. Calculation of the channel resolvent $G_1$. \\
5. Solution of the resulting matrix equation by iteration using the
Pade-approximant
technique. \\

The runtimes for the steps 1 and 4 are negligible in comparison with
the total running time,  so that we can leave these steps for
sequential execution on the CPU.
The main computational effort (in the CPU realization)  is
spent just on the calculation of the matrix elements of ${\mathbb
P}^0$-matrix (i.e. the step 3) which are calculated independently
of each other. So this step is ideal for parallelization and we
primarily parallelized only the corresponding part of the computer
code. However, since the matrix ${\mathbb P}^0$ is very sparse, in
order to reach the high efficiency of its parallel computation, the
preliminary selection of its nonzero elements should be carried out
(step 2). The execution of the 5th step --- iteration solution
of the matrix equation --- can certainly be accelerated using linear
algebra routines implemented on the GPU, but as its execution time
takes no more than 20\% of the total time for solving the whole
problem we did not optimize this step in the present study. Thus, 
in this calculation we have only parallelized 
the time-consuming steps 2 and 3 for the GPU. 

\subsection{Comparison of the GPU and CPU realizations}

In this section we compare the runtimes for the CPU and GPU realizations
of realistic $nd$ elastic scattering calculations.  For our GPU
calculations we used an ordinary NVIDIA GTX-670 video card, which is
not specialized for general-purpose computing.

First we compare the CPU and GPU realizations of the above
algorithm for case of the $S$-wave MT I-III $NN$ interaction . This
example clearly demonstrates the advantage of the GPU
acceleration.

\begin{figure}[h!] \epsfig{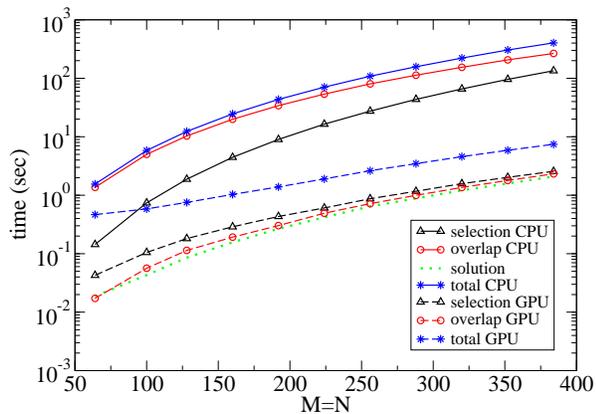}
\caption{(Color online)  The dependence of the computing time on CPU (solid
lines) and
 GPU (dashed lines) (for the full solution and separate steps) on
dimension of the basis $M=N$ for the $nd$ scattering problem with
$S$-wave $NN$ interaction.  \label{time_GPU}}
\end{figure}

Fig.~\ref{time_GPU} shows the dependence of the CPU- and
GPU-computing times for the above steps 2, 3,  5 and the
complete solution for dimensions $M$ and $N$ of the  WP bases chosen
along both Jacobi variables. We take  for simplicity $M=N$ in all
of our tests so that, the total dimension of the matrix kernel is equal
to $M^2$.

It is clear from Fig.~\ref{time_GPU} that the complete solution of
$nd$ scattering problem in our approach for a basis of the large
dimension\footnote{We take here a basis of this dimension just for a
numerical test. To solve accurately the initial physical problem,
the basis with $M=N=100$ is quite enough.} $M=400$ takes only ca.
7~sec on a serial PC using GPU! This ultra-fast solution
handles a huge matrix  $160000\times 160000$ including the
calculation of the ca. $M^4/100=256$ millions nonzero elements of
matrix ${\mathbb P}^0$ where each matrix element is reduced to a
integral of a rather complicated algebraic function. The integrals
are calculated numerically with a 48-grid-point Gaussian quadrature.

\begin{figure}[h!]  \epsfig{file=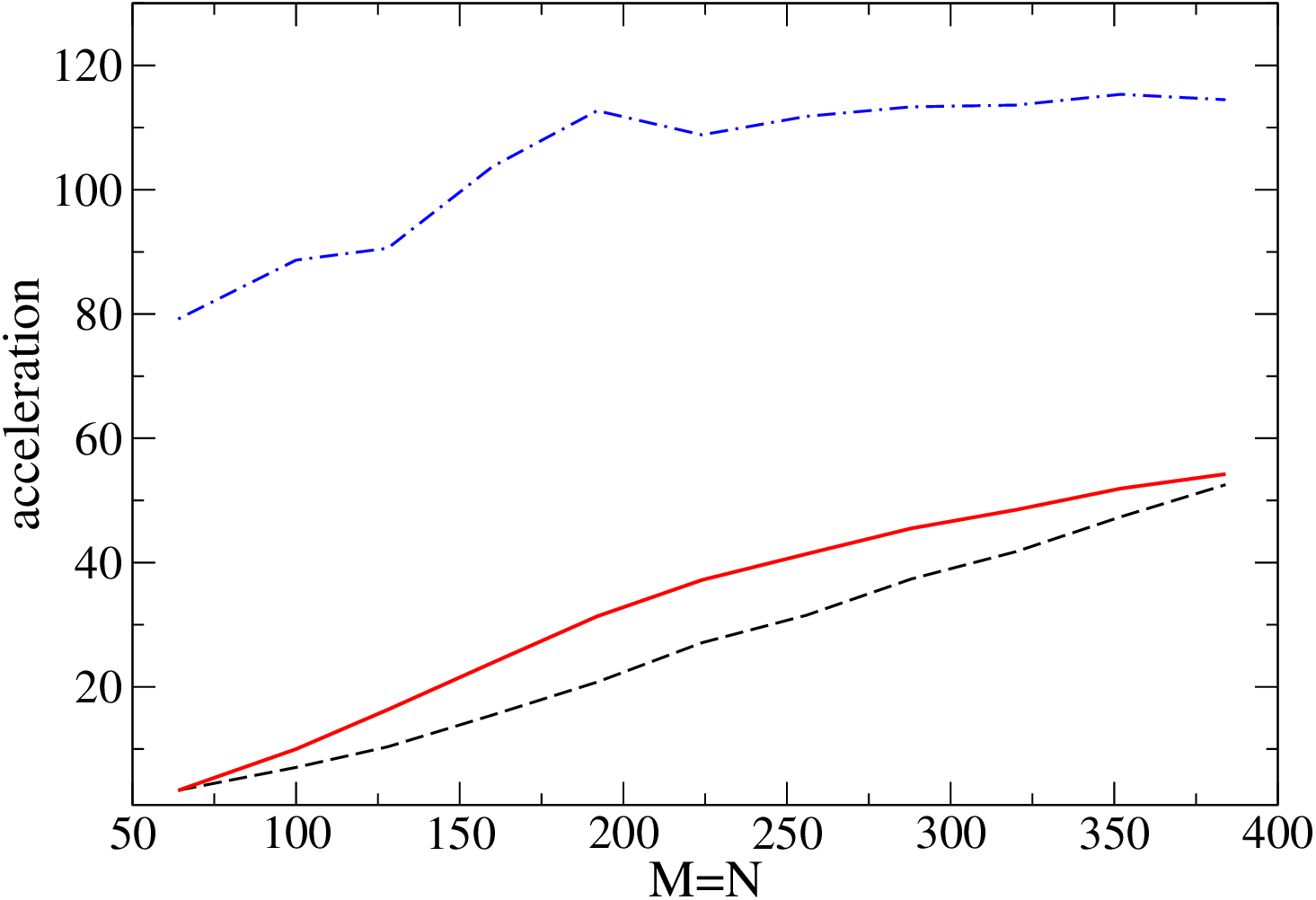,width=0.9\columnwidth}
\caption{(Color online)  The dependence of the  GPU-acceleration
ratio $\eta$ on the dimension of the basis $M=N$ in the calculation of
the $nd$ scattering problem with a $S$-wave $NN$
 interaction: dashed line shows the acceleration for the step 2 (selection),
 dot-dashed line --- for the step 3 (calculation of ${\mathbb P}^0$),
 solid line ---  acceleration for the complete solution.}
\label{accel_GPU}
\end{figure}

From Fig.~\ref{time_GPU} it is seen that all of these 256 million 
integrals can be computed on GPU in just 2.3 seconds compared to 255
seconds on the CPU. This demonstrates the real very high speed of GPU
computations for the case discussed here.

In Fig.~\ref{accel_GPU} we present the dependence of the GPU
acceleration ratio $\eta=t(CPU)/t(GPU)$ on the dimension of the basis
$M=N$ for the solution of the $S$-wave Faddeev problem. The total
acceleration for the complete solution varies from 10 to 50 times
depending on the basis dimension, while the time for calculating the
nonzero elements of the matrix ${\mathbb P}^0$ (the step 3) which
takes the main part of the CPU-computing time is reduced by a factor
of more than 100.

\begin{figure}[h!]
 \epsfig{file=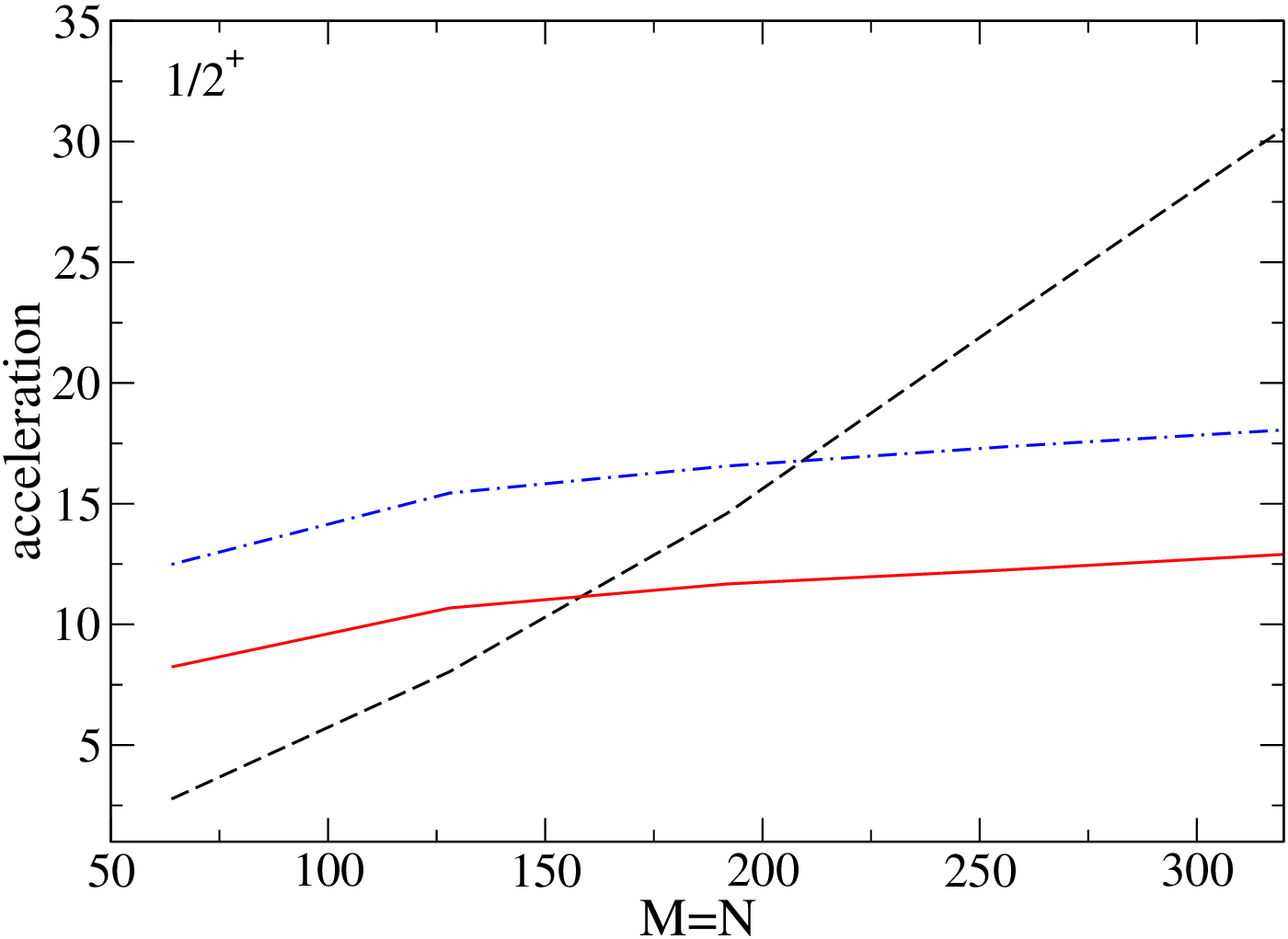,width=0.85\columnwidth}
\caption{(Color online)  The dependence of the GPU acceleration
ratio $\eta$ on the dimension of the basis $M=N$ for the realistic $nd$
scattering problem for $J=\frac12^+$: dashed line shows the
acceleration for the step 2 (selection),
 dot-dashed line --- for the step 3 (calculation of ${\mathbb P}^0$),
 solid line -- the  acceleration for the complete solution.
 \label{time_GPU_Nijm}}
\end{figure}

We now turn to the case of the realistic $nd$  scattering problem with
the Nijmegen $NN$ potential.  Taking into account higher partial
waves leads to the system of coupled two-dimensional integral
Faddeev equations.  Now the calculation of each matrix element of
${\mathbb P}^0$ for partial waves with nonzero angular momenta
includes several tens of double numerical integrals with some
trigonometric functions and uses a large set of algebraic
coefficients $g^{l_1l'_1k}_{\ga\ga'}$ (\ref{gal}).  The proper
parallelization for this
calculation with pre-selection
of the nonzero matrix elements leads again to rather fast algorithm
realized on GPU.  Fig.~\ref{time_GPU_Nijm} demonstrates the GPU
acceleration ratio $\eta=t(CPU)/t(GPU)$ for the complete solution
(solid line) and for the steps 2 (dashed line) and 3 (dot-dashed)
as a function of the basis dimension $M=N$ for the solution of the 
18-channel Faddeev equations for the $J=\frac12^+$ 
partial $nd$ elastic amplitude.

It is evident from the results presented that the passing from CPU-
to GPU-realization {\em on the same PC} allows one to obtain a
significant acceleration of whole three-body calculation by an
order of magnitude or higher. It is clear also that the use of a
more powerful specialized graphics processor, like the Tesla K40,  would
lead even to a considerably higher acceleration of  calculations.
However, it should be emphasized that the total acceleration which can
be achieved by using GPU depends crucially on the method used, the
numerical scheme and  parallelization implementation.

\section{Conclusion}
In conclusion, let's summarize  the most important points of the
proposed new general approach for the solution of few-body scattering
problems.

1. First, we rewrite the initial Faddeev equations, which include
an off-shell $t$-matrix, in a fully equivalent form which
incorporates the product of the channel resolvent and the
interaction operator. This allows us to simplify the required two-body
input and also to fix all of the energy singularities in the channel
resolvent operator.  We note here that the  four- and more
particle integral Faddeev-Yakubovsky equations have, in principle,
some similar structure as compared to three-body Faddeev equations
(surely being more complicated and having a higher dimension).
Therefore the discrete WP technique outlined in this paper
can be employed for solving these general equations as well.

2. After that, we project out the transition and interaction
operators onto a discrete wave-packet basis and employ
a specific analytical form for  one- and multi-channel
resolvent operators. This discrete form is extremely convenient for
few-body calculations because we get a fixed matrix form (with
fully regular matrix elements) for the scattering equations (of
Lippmann--Schwinger, Faddeev or Faddeev-Yakubovsky types). The most
important improvement in such a WP projection is the fact that  we can
replace the permutation operators by  fixed matrices, thus
completely avoiding the numerous and time consuming  interpolations
at every iteration step. Moreover, this matrix is energy- and
interaction-independent and, being calculated once, can be used to
solve various scattering problems with different potentials and at
many energies.

These two steps lead  to a rather effective numerical scheme which
is realized on a standard  PC for realistic $3N$ scattering 
calculations.

3. At last, we perform a proper parallelization of the solution
and eventually we employ the ultra-fast GPU technique to make very
effective parallel calculations on a serial PC. This GPU realization
for realistic scattering problems reduces the
computational time by at least one order of magnitude  (while
for separate parts of the numerical scheme the total acceleration
could reach two orders).

All of the above points open a new way 
for doing extensive many-body scattering calculations via continuum
discretization e.g. in quantum chemistry,  nuclear reaction
theory,  solid state theory etc..

There is little doubt that a similar (but surely more tedious)
approach can also be applied to other scattering problems, e.g. to
solve the relativistic Faddeev equations, Bethe--Salpeter equations
etc.

 {\bf Acknowledgments} The authors thank Dr. A.V.~Boreskov for discussions
 of  problems associated with a use of the GPU. The authors are also deeply
 grateful to Prof. W. Polyzou for careful reading of the manuscript and valuable
 comments. This work has been supported
 partially by the Russian Foundation for Basic Research,
 grants Nos. 12-02-00908 and 13-02-00399.

\appendix

\section{Calculation of permutation matrix elements.}
The  kernel of the permutation operator in momentum space has the
form \cite{gloeckle}
\begin{equation}
\langle pq\gamma|P|p'q'\gamma'\rangle=\int_{-1}^1 dx
\frac{\delta(p'-\pi_1)\delta(p-\pi_2)}{(p')^{l'+2}p^{l+2}}
G_{\gamma\gamma'}(q,q',x),\label{P_kern}
\end{equation}
where the multi-indices $\gamma$ and $\gamma'$ include all possible
spin-angular quantum numbers for the three-body states (as they are
detailed in the Sec.IV , i.e. $\gamma=l,s,j,\lambda,I,J,\pi,t,T$)
and the following notations
\begin{equation}
 \pi_1=\sqrt{q^2+\quat(q')^2+qq'x},\
\pi_2=\sqrt{\quat q^2+(q')^2+qq'x},
\end{equation}
are used. The spin-angular coefficients  $G$ can be found from the
formula
\begin{equation}
G_{\ga\ga'}(q,q',x)=\sum_{l_1,l'_1,k}q^{l_2+l'_2}(q')^{l_1+l'_1}P_k(x)g^{l_1l'_1k}_{\ga\ga'},\
   \begin{array}{c}
   l_1+l_2=l,\\ l'_1+l'_2=l',\\
   \end{array}
   \label{ggaga}
\end{equation}
where $l$ and $l'$ are two-body subsystem orbital momenta and the 
indices $l_1,l_2,k,l'_1,l'_2$ arise from  intermediate triangle
sums. The sum in (\ref{ggaga}) runs according to the triangle rules
in partial coefficient $g^{l_1l'_1k}_{\ga\ga'}$ which have the
following explicit form \cite{gloeckle}:
\begin{align}
g_{\ga\ga'}^{l_1l'_1k}=-\sqrt{\hat{l}\hat{s}\hat{j}\hat{t}
\hat{\lambda}\hat{I}\hat{l'}\hat{s'}\hat{j'}\hat{t'}\hat{\lambda'}
\hat{I'}}
\left\{\begin{array}{ccc}\frac{1}{2}&\frac{1}{2}&t'\\
\frac{1}{2}&T&t\end{array}\right\}
 \sum_{LS}(\hat{L}\hat{S})\times\nonumber\\
\left\{\begin{array}{ccc}\frac{1}{2}&\frac{1}{2}&s'\\
\frac{1}{2}&S&s\end{array}\right\}
\left\{\begin{array}{ccc}l&s&j\\
\lambda&\frac{1}{2}&I\\L&S&J
\end{array}\right\}
\left\{\begin{array}{ccc}l'&s'&j'\\ \lambda'&\frac{1}{2}&I'\\L&S&J
\end{array}\right\} \times\nonumber\\
 \hat{k}\left(\frac{1}{2}\right)^{l'_2+l_1}
\sqrt{\frac{(2l+1)!}{(2l_1)!(2l_2)!}}
\sqrt{\frac{(2l'+1)!}{(2l'_1)!(2l'_2)!}} \times\nonumber\\
 \sum_{ff'} \left\{\begin{array}{ccc}l'_1&l'_2&l'\\
\lambda'&L&f'\end{array}\right\}
\left\{\begin{array}{ccc}l_2&l_1&l\\
\lambda&L&f\end{array}\right\}
\langle l'_20\lambda' 0|f'0\rangle \langle l_10\lambda 0|f0\rangle\times \nonumber \\
\left\{\begin{array}{ccc}f'&l'_1&L\\
f&l_2&k\end{array}\right\} \langle k0l'_10|f0\rangle \langle
k0l_20|f'0\rangle,\label{gal}
\end{align}
where the summation is done over all allowed intermediate quantum
numbers.
 Projection of free three-body wave-functions  $|p,q\rangle$ (plane waves) onto the WP states can
be found (for the case of unit weights $f(p)$ and $\bar{f}(q)$) from
the formula \cite{nd2} :
\begin{equation}
\langle
p,q|\mfp_i\mfq_j\rangle=\frac{1}{\sqrt{d_i\bar{d}_j}}\frac{\theta(p\in\MD_i
)\theta(q\in\BMD_j)}{pq}.
\end{equation}
Here $d_i$, $\bar{d}_j$ are the widths of momentum intervals and function $\theta$ is analog of the Heaviside step-function:
$\theta(p\in\MD_i)=1$ if $p$ belongs to the  interval $\MD_i$ and
$\theta(p\in\MD_i)=0$ otherwise.

To obtain the permutation matrix elements in the lattice basis
(\ref{perm}), one should integrate kernel (\ref{P_kern}) over the
lattice cells $\MD_{ij}$ and $\MD_{i'j'}$.
These matrix elements are defined by the following
integrals:
\begin{eqnarray}
 [{\mathbb P}_0]_{ij,i'j'}^{\ga\ga'}\equiv
\langle
X_{ij}^\ga|P|X_{i'j'}^{\ga'}\rangle=\sum_{kl_1l'_1}\frac{g_{\ga\ga'}^{kl_1l'_1}}
{\sqrt{d_i\bar{d}_j d_{i'}\bar{d}_{j'}}}\times\nonumber\\
\int_{\MD_{ij}} dpdq \int_{\MD_{i'j'}}dp'dq' \int_{-1}^1dx
\frac{q^{l_2+l'_2+1}(q')^{l_1+l'_1+1}}{(p')^{l'+1}p^{l+1}}\times\nonumber\\
\delta(p'-\pi_1)\delta(p-\pi_2)P_k(x).\qquad
\end{eqnarray}
Using the Dirac delta-functions, the integrals over $p$ and $p'$ are
evaluated analytically:
\begin{eqnarray}
 [{\mathbb
P}_0]_{ij,i'j'}^{\ga\ga'}=\sum_{kl_1l'_1}\frac{g_{\ga\ga'}}
{\sqrt{d_i\bar{d}_j d_{i'}\bar{d}_{j'}}}\int_{\BMD_{j}} dq
\int_{\BMD_{j'}}dq' \int_{-1}^1dx\nonumber\times\\
\frac{q^{l_2+l'_2+1}(q')^{l_1+l'_1+1}}{(\pi_1)^{l'+1}(\pi_2)^{l+1}}P_k(x)\theta(\pi_1\in\MD_{i'})
\theta(\pi_2\in\MD_i).\qquad
\end{eqnarray}
Further, the residual integrals can be evaluated in polar
coordinates:
\begin{equation}
q=Q\cos\phi,\quad q'=Q\sin\phi.
\end{equation}
Then the matrix element takes the form:
\begin{equation}
[{\mathbb
P}_0]_{ij,i'j'}^{\ga\ga'}=\sum_{kl_1l'_1}\frac{g_{\ga\ga'}}
{\sqrt{d_i\bar{d}_j d_{i'}\bar{d}_{j'}}} I,
\end{equation}
where
\begin{eqnarray} I=\int_{-1}^1dx
P_k(x)\int_{\phi_1}^{\phi_2}d\phi
\frac{(\cos\phi)^{l_2+l'_2+1}(\sin\phi)^{l_1+l'_1+1}}{(\zeta_1)^{l'+1}(\zeta_2)^{l+1}}
\times\nonumber\\
\int_{Q_1(\phi)}^{Q_2(\phi)}QdQ\
\theta\left(Q\in\frac{\MD_{i'}}{\zeta_1}\right)
\theta\left(Q\in\frac{\MD_{i}}{\zeta_2}\right),\qquad
\end{eqnarray}
and the following notations are employed:
\begin{eqnarray}
\zeta_1=\sqrt{\cos^2\phi+\quat\sin^2\phi+x\cos\phi\sin\phi} ,\nonumber\\
\zeta_2=\sqrt{\quat\cos^2\phi+\sin^2\phi+x\cos\phi\sin\phi}.
\end{eqnarray}
The boundaries of the integral area in $(Q,\phi)$-plane are defined
by transformation of the rectangle in the $(q,q')$-plane, so that:
\begin{equation}
\begin{array}{ll}
\phi_1=\arctan\left(\frac{q'_{j'-1}}{q_{j}}\right),&
Q_1(\phi)=\max\left(\frac{q_{j-1}}{\cos\phi},\frac{q'_{j'-1}}{\sin\phi}\right),\\
\phi_2=\arctan\left(\frac{q'_{j'}}{q_{j-1}}\right),&
Q_2(\phi)=\min\left(\frac{q_{j}}{\cos\phi},\frac{q'_{j'}}{\sin\phi}\right).\\
\end{array}
\end{equation}
Evaluating the $Q$-integral analytically, one gets the expression
for the integral $I$:
\begin{eqnarray}
I=\int_{-1}^1dx P_k(x)\int_{\phi_1}^{\phi_2}d\phi
F_{ij,i'j'}(\phi,x) \times\nonumber\\
\frac{(\cos\phi)^{l_2+l'_2+1}(\sin\phi)^{l_1+l'_1+1}}{(\zeta_1)^{l'+1}(\zeta_2)^{l+1}}
,
\end{eqnarray}
where the following function is introduced:
\begin{eqnarray}
F_{ij,i'j'}(\phi,x)=\half\left[\min\left(\frac{q_{j}}{\cos\phi},\frac{q'_{j'}}{\sin\phi},
\frac{p'_{i'}}{\zeta_1},\frac{p_{i}}{\zeta_2}\right)\right]^2
-\nonumber\\
\half\left[\max\left(\frac{q_{j-1}}{\cos\phi},\frac{q'_{j'-1}}{\sin\phi},
\frac{p'_{i'-1}}{\zeta_1},\frac{p_{i-1}}{\zeta_2}\right)\right]^2.\qquad
\qquad
\end{eqnarray}
Finally, one gets the eventual formula for the permutation matrix
\begin{eqnarray}
[{\mathbb
P}_0]_{ij,i'j'}^{\ga\ga'}=\sum_{kl_1l'_1}\frac{g_{\ga\ga'}^{kl_1l'_1}}
{\sqrt{d_i\bar{d}_j d_{i'}\bar{d}_{j'}}}\int_{-1}^1dx
P_k(x)\times\nonumber\\
\int_{\phi_1}^{\phi_2}d\phi F_{ij,i'j'}(\phi,x)
\frac{(\cos\phi)^{l_2+l'_2+1}(\sin\phi)^{l_1+l'_1+1}}{(\zeta_1)^{l'+1}(\zeta_2)^{l+1}}.
\end{eqnarray}
In practical treatments these integrals are evaluated numerically.

 \end{document}